# COULOMB INTERACTION AS THE SOURCE

# OF MUSCLE FORCE

E.V. Rosenfeld

Russian Academy of Sciences, Ural Branch
Institute of Metal Physics
Kovalevskaya str. 18, 620219 Ekaterinburg, Russia

e-mail: rosenfeld@imp.uran.ru

Tel: 7(343)3757970; 7(343)3783570
Fax: 7(343)3745244

For displacement of individual fragments of large molecule relative to each other to arise and work to be performed, force must be born *inside* protein. W*hat kind of interaction generates this force*? Models based on Huxley's 1957 theory ascertain relations between chemical reactions rate constants and energies of crossbridge conformations. Nevertheless, understand in the framework of thermodynamics how myosin motor works in principle is impossible: it is smoothly heated device cyclically producing mechanical work (second law). Furthermore, in every working cycle myosin head captures and splits a *single* ATP molecule. Hence, ordinary dynamic laws rather than stochastic laws govern this process. The simple mechanism of chemomechanical transduction is proposed in this work. The two products of *ATP* hydrolysis, *ADP* and $P_i$, have charges of the same sense and Coulomb interaction of these charges after hydrolysis produces the force pushing backdoor and rotating converter domain. The velocity of filaments sliding becomes the principal parameter of the model and new mechanism of indirect interaction between the cross-bridges radically different from one suggested by Huxley and Simmons in 1971 appears. The working stroke duration is inversely proportional to the sliding velocity now. Therefore Hill equation appears and the parameter values obtained are in reasonable agreement with experiment.

## I. INTRODUCTION

The overwhelming majority of human-made devices that perform mechanical work are driven by solely two kinds of forces. The first case is the force of pressure onto a certain surface of fast-moving molecules, and the second, ponderomotive forces of interaction of electric currents and/or magnetized cores. However, inside a nature-produced engine – muscle the temperature distribution is virtually homogeneous and no appreciable flows of mass or charge are observed in it. Therefore, it is absolutely evident that a muscle is driven by forces of another type, and the vital question to be asked while investigating the mechanism of muscle performance is *what kind of interaction is the source of the muscle force*.

The necessity for formulating the problem in such a way becomes evident in the light of astonishing progresses in studies into the structural mechanism of the muscle contraction (Cooke, 2004; Geeves and Holmes, 1999; Gordon, Homsher and Regner, 2000). Anyhow, the progress in comprehension of the mechanical structure of any device, no matter a cross-bridge or an alien space ship, has to be followed by progress in understanding of the foundational principle of this mechanism operation. The structure of any well-designed device should first correspond the best to the physical principle of its operation and further meet the requirements for practicability, efficiency, etc. Consequently, having clarified what elements constitute the cross-bridge and wishing to go further and make it clear why it has this very structure, we have to ascertain first what kind of force sets the cross-bridge into motion, i.e. what kind of force is the cause of its conformation.

An exclusive complexity of the processes conditioning and accompanying the protein conformation is indubitable. However, we'll fail to understand the principle of myosin motor performance until we are able though in a rough and simple approximation to understand the essence of these processes. After all, protein conformation is a mere change of its shape, i.e. displacement of individual fragments of a large molecule relative to each other. Consequently, for this displacement to arise and, moreover, in doing so, for mechanical work to be performed over external objects, forces must be born *inside* protein that put its separate parts into motion. Exact calculation of these forces is a problem naturally unsolvable so far because of complexity of protein molecules. However, the ascertainment of the nature of these forces and simple numerical estimations seem far simpler and thus quite resolvable problem.

There exist many works in which different models of myosin motors working due to hydrolysis of the adenosine triphosphate molecule (*ATP*) are constructed. However, the majority of them from the one become classical long ago A.F Huxley's 1957 theory (Huxley, 1957) to comparatively latest (Baker et al, 1999; Duke, 1999; Pate and Cooke, 1989; Piazzesi and Lombardi, 1995; Smith and Geeves, 1995) are based on one and the same idea – the ascertainment of relation between the rate constants of chemical reactions and free energies of different cross-bridge conformations. As the least indivisible structural unit in this case, the myosin head is taken, which periodically changes its shape and attaches to and detaches from the actin filament. Since these changes are unambiguously ascribed to individual stages of the *ATP* hydrolysis (Lynn-Taylor scheme (Lynn and Taylor 1971)), such an approach presents the opportunity to determine the ratio of time intervals corresponding to the myosin head occurrence in different conformations. The importance of such researches can hardly be overestimated, but it is necessary to underline that to understand in the framework of classical chemical thermodynamics how myosin motor works in principle is impossible.

Indeed, in equilibrium both courses of the reaction are equivalent already because it is defined solely by the ratio of *scalars*: free energies and temperature. To describe a unidirectional process the theory should operate with *vectors*. Say, in the frame of non-equilibrium thermodynamics upon external changes in the component concentrations, there arise appropriate generalized forces and generalized flows. It is clear that in a contracting muscle both the preferable reaction course and the degree of deviation from equilibrium are controlled by the velocity of filament sliding; hence, this velocity must indispensably enter a coherent theory.

Second, myosin head is a device cyclically producing mechanical work under conditions when all its parts have virtually the same temperature. Therefore, it is clear that in the framework of pure thermodynamics, one of the most fundamental concepts of which is its second law, the device performance is impossible to understand. It is just the reason for this very process of the force emergence to not be discussed in most theories; it is commonly postulated that a cross-bridge attached to actin generates a force. However, it requires mentioning herein that there exist quite a number of works, published in different forms, in which the problem on the nature of the force moving the myosin motor is distinctly formulated and in a way solved. Let me cite one of these works (Oster and Wang, 2003). "The basic physical principle that governs the operation of all protein motors … is this: molecular motors generate mechanical forces by using intermolecular binding energy to capture "favorable" Brownian motions". A brilliant analysis of the performance of such a motor in The Feynman lectures on physics (Feynman, 1963) shows, however, that such an explanation of the force nature is acceptable only if the second law of thermodynamics is considered inapplicable to animated organisms.

Finally, a view beyond the scope of thermodynamics in the investigation of myosin motor is essential already because in every working cycle of this device we deal with capturing and splitting of a *single ATP* molecule. Hence, it immediately follows that this process is governed by ordinary dynamics laws rather than stochastic laws for the motion of macroscopic ensembles of molecules. For the same reason, trials to relate the emergence of a force driving myosin head to the changes in the entropy of a single molecule can not be considered substantiated: application of the concept "entropy" to a single molecule is meaningless.

So, to understand the nature of the force driving myosin motor a simple mechanical model is required. Naturally, with a strict approach to the problem it is necessary first to use quantum rather than classical mechanics, and second, take into account fluctuations. However, in a simplest version of analysis of the myosin motor performance, which is presented in what follows, only classical mechanics methods are applied.

Let me clarify the above using a trivial but widely used example. What can we learn about the vehicle system if we restrict ourselves, when investigating it, to only kinematical and energy aspects of the problem? We apparently come to three main conclusions:

1. A car moves owing to the chemical energy conversion to mechanical one that takes place in an engine with certain efficiency.

2. The reason for the force that drives a car to arise consists in conformational changes that occur in the block of cylinders.

3. Propulsive effort and velocity of motion are unambiguously related to separate step durations in this conformation cycle.

It is the same thing or near so what we know about myosin motor operation. Formally, these statements are quite valid, but it can not help us clarify, for example, whether we deal with steam engine or combustion engine. Certainly, if we know that steam engine ejects steam, whereas combustion engine, exhaust, to tell one engine from another is of no difficulty. Yet, if we, just as in the case of myosin motor handle with an absolutely unknown device which we fail to look into, we have a classical case of black box. In this situation, models of its internal arrangement are required to be constructed when trying to find the one which most precisely reproduces the main results of the object performance. In particular, when considering an apparatus that performs mechanical work $W$, the most important is, in accordance with the first thermodynamics law

$$\Delta U = -\Delta Q - \Delta W \qquad \text{(Eq1)}$$

 the correct reconstitution of heat release $Q$ and the force versus displacement (or velocity) dependences. In the work presented, the nature of the driving force for muscle is postulated from general reasoning (see the next section) and the findings gained from this postulate are compared with the results of just such-kind experiments excellent reviewed in (Bendall, 1969; Woledge, Curtin and Homsher, 1985).

## II. MECHANICAL WORK PERFORMED DURING CHEMICAL REACTION

As to the force that causes a muscle to move, it arises during the $ATP$ splitting and that is why the question on the origin of this force is directly related to the issue why it is $ATP$ that is the energy source for muscular contraction. In other words, a problem should be set on what it is in the structure of the $ATP$ molecule that provides a capacity to generate force and perform mechanical work in the course of hydrolysis.

The only interaction that completely controls the course of chemical reactions is electromagnetic interaction. The magnetic part of it, i.e. the energy of direct magnetic interaction of reacting molecules always is virtually negligibly small. Consequently, the only force that can perform mechanical work during a chemical reaction, i.e., upon rearrangement of atomic electron shells, is the Coulomb force. Therefore, all that is set forth below is based on the following definition, which can be treated as the main postulate of this paper as well.

*Direct conversion of chemical energy to mechanical one is utilization of the mechanical work performed by the Coulomb forces when rearranging the electrical charges in the course of chemical reaction.*

Thus, to understand the mechanism of the force generation, one should ascertain what charges shift upon the *ATP* cleavage and in which directions. If to take into account that in the cell-milieu the *ATP* molecule is a tetraanion which upon the $ATP \rightarrow ADP + P_i$ hydrolysis decomposes into like charged fragments  adenosine diphosphate $ADP^{2-}$ and inorganic phosphate $P_i^{2-}$ (Bendall, 1969; Bohinski, 1983), the Coulomb repulsion of these pieces can quite naturally be considered as the force propelling myosin motor..

This principle is quite similar to one underlying the firearms (or combustion engine) performance. In the course of a shot the cartridge breaks up into the bullet and the cartridge-box (compare *ATP* cleavage). The force of hot powder-gas pressure (compare Coulomb force) acts on these pieces in the opposite directions performing mechanical work. The energy of the bullet is the more, the longer distance it passes inside a trunk, and when the bullet leaves the trunk, the residuary powder-gas energy dissipates. It is this analogy that one would have in view hereinafter.

I would let myself quote here a short fragment from the book (Woledge, Curtin and Homsher, 1985), where the correspondence of different stages of *ATP* hydrolysis to those of the cross-bridge working cycle (Lynn-Taylor scheme, see here Fig. 1, the copy of Fig. 1.14 from (Woledge, Curtin and Homsher, 1985)) is established. "The working stroke of the cross-bridge might be tentatively identified with the release of products from the actomyosin-product complex (step 4 in Fig.1) and the recovery stroke, with the splitting step (step 2). It is worth noting that in this scheme *ATP* is split on myosin alone, not when actin and myosin are combined. Also, external work is not done at the time of *ATP* splitting; *ATP* splitting puts myosin into a state in which it can do work."

This viewpoint is apparently based on the standard assumption that the potential energy of *ATP* is concentrated in the deformed *P-O-P*-bond and releases in the course of its breaking. As a result, one has to suppose that the released energy becomes somehow stored, that is, a certain myosin "spring" becomes stretched out ("*ATP* splitting puts myosin into a state in which it can do work"). Later, *ADP* and $P_i$ remain localized and the "spring" stretched out as long as the actomyosin complex restores again after step 3. Then *ADP* and $P_i$ are released whereas "myosin spring" relaxes, moving filaments and executing work.

**(A)**  **(B)**

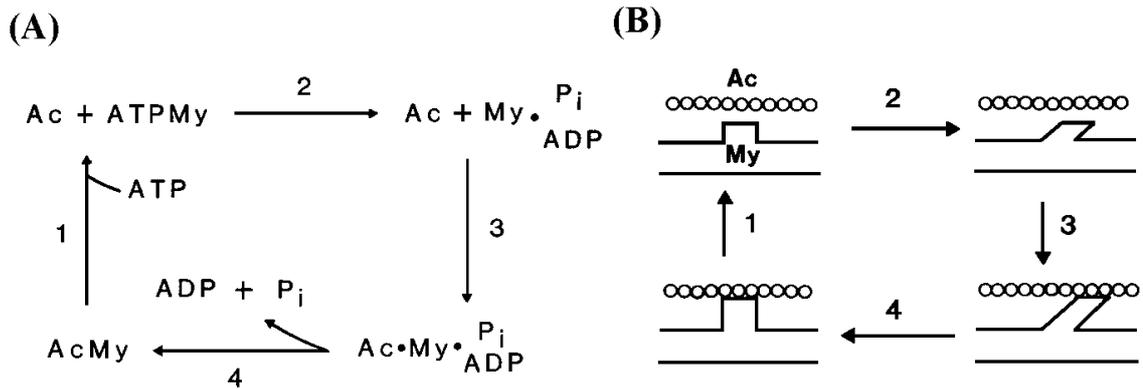

**Figure 1** A simple analogy between *ATP* splitting by actomyosin in solution (A) and a cross-bridge cycle (B). Ac, Actin, My, Myosin. The copy of Fig. 1.14 from (Woledge, Curtin and Homsher, 1985)

Under assumption that the release of *ADP* and $P_i$ occurs at the very beginning of the working stroke (see Fig.1), the speculate concerning the existence of a "myosin spring" is the only way reasonable. However, it is known nowadays that $P_i$ releases prior to *ADP*, and it can leave the *ATP*-binding-center only after the switch-2, i.e. backdoor, opens (Geeves, and Holmes 1999). This means that the two likely charged *ATP* fragments remain closely packed after the *P-O-P*-bond breaking in the course of the working stroke and are similar in themselves to the stressed spring. Hence, we obtain much simpler scheme, supposing that it is just the Coulomb force acting on $P_i$ that pushes switch-2, thus performing work and making the way out (opening the door) into the intracellular liquid.

In the framework of this approach, many principle questions are sure to remain open, for example:

1. What is the nature of the close actin-myosin link?
2. Why does this link break when *ATP* molecule connects to the cross-bridge?
3. How does the *ATP* hydrolyses occur?
4. Why does not cross-bridge begin its working stroke until an actomyosin complex restores?

and so on. However, these very questions remain open in other cross-bridge models as well. Besides, being extremely complex and important by themselves, these questions are of solely indirect concern with the central problem of this work.

So, using this approach, we do not miss anything and at the same time we acquire pretty much. We obtain a connection between the force and distance separating *ADP* and $P_i$ that is determined by the Coulomb law and is discussed in the next section. Besides, we obtain also the connection between the working stroke duration and the velocity of relative filaments displacement. Hill equation wind up the straightforward consequence of this connection, and the question is discussed in the next sections too. Lastly, there is no more need to complicate matters and consider some special high-energy "myosin state in which it can do work".

I certainly don't mean that the *ATP* hydrolysis does not result in any changes in the structure of myosin head. These changes naturally take place and it is just these changes that give rise to recovering stroke and changes in the affinity of the head for the actin. It is quite natural to assume that it takes some part of energy being released in the course of *ATP* hydrolysis (see also section Discussion). Nevertheless, the main part of this energy, which is then spent on mechanical work and heat release, does not need storing in myosin since it is already stored in the form of energy of electrostatic interaction of the hydrolysis products.

In this connection the recent work (Lampinen and Noponen, 2005) is worth mentioning. In the work the electric-dipole theory of storing and transforming the *ATP* chemical energy in actomyosin molecular motor is presented. The authors of this work suppose that cross-bridge conformation is its respond to the change in the dipole electric field of *ATP* molecule, which occurs when phosphate tail shortens. However, the dipole field represents the major part of the electric field of some system only in the case when the system as a whole does not have electrical charge. Otherwise, the common Coulomb interaction becomes forefront, and this interaction is taken into account here for charged *ATP* molecule, which apparently is the case of the cell-milieu.

### III. PHENOMENOLOGICAL MODEL

The Coulomb force acting on the two charged spherules and Coulomb energy stored by them can be estimated as usual

$$F = \frac{1}{4\pi\varepsilon_0} \frac{q_1 q_2}{\varepsilon r^2} \approx 10^{10} \frac{q_1 q_2}{\varepsilon r^2} \text{ N}, \quad E = \frac{1}{4\pi\varepsilon_0} \frac{q_1 q_2}{\varepsilon r} \approx 10^{10} \frac{q_1 q_2}{\varepsilon r} \text{ J}. \tag{Eq2}$$

Here $q_1$ and $q_2$ are the charges of the spherules, $r$ is the distance between its centers, and $\varepsilon$ is the dielectric permeability of the medium. Let consider two such spherules with the charges of the same sense closely packed in the cylinder with piston; see Fig.2 where the initial positions of the spherule 2, the piston and the lever are shown by dashed lines. Then, spherule 2 displaces, being pushed by Coulomb force, and moves the piston rod connected to the lever. The lever in its turn pushes attached to it rope moving with the velocity *v*. The work *W* is evidently performed by this system until spherule 2 displacement increases up to *D* and it drifts out of the cylinder. We denote the duration of this "working stroke" as $\tau_w$, and the displacement of the rope during this lapse, S. Next; the lever is detached from the rope and returned to the initial position, the cylinder is recharged, the lever becomes connected to the rope again, and the process repeats with a period T. It is obvious that $T = \tau_r + \tau_w$ where $\tau_r$ is the recharging (recovering) time independent of the rope velocity $v$.

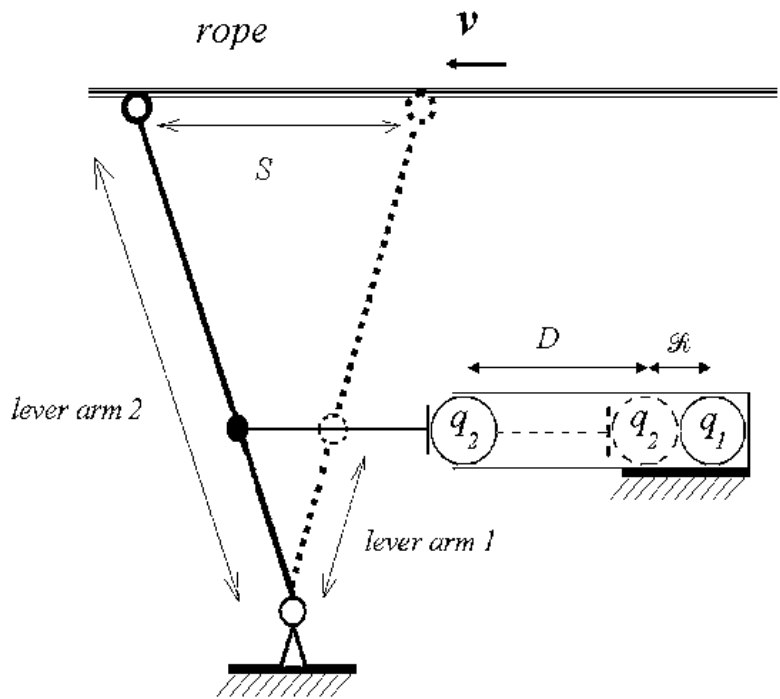

**Figure 2**.   Key scheme of an engine moving a rope. In the initial position (dashed lines) the distance between the centers of the spherules $q_1$ and $q_2$ is equal to $\mathcal{R}$. Then the repulsive Coulomb force displaces $q_2$ and the distance increases up to $\mathcal{R}+D$ while the corresponding rope displacement is S. The lever arms ratio $\lambda$=(*lever arm 2*): (*lever arm 1*)= S/D, see (Eq3).

The particular type of the transmission "gear" (lever in Fig.2) is not decisive for us now; assume only that

I.   it functions without losses,

II.   independently of the design of the transmission gear, the shift $D$ at its inlet is proportional to the shift $S$ at its outlet

$$S = \lambda D. \qquad (Eq3)$$

Hence, the distance $D$, which the spherule 2, being pressed to the piston, passes under the Coulomb force, is $\lambda$ times as low as the relative rope displacement $S$ for the same time, whereas the force acting on the charges is $\lambda$ times as high as that acting on the rope. In this section, we concretize neither the real values of $q_1$, $q_2$, $\lambda$, nor the properties of the medium inside which the charges move ($\varepsilon$). Nevertheless, we can try to employ dependence (Eq2) to construct a simple phenomenological model in which the starting distance between the spherules and the starting force proportional to the product of their charges divided by dielectric permeability ($q_1 q_2/\varepsilon$) are taken as the basic parameters.

Suppose that the centers of spherules 1 and 2 at a moment $t = 0$ are localized at a distance $\mathcal{R}$ and the Coulomb force acting on the charges is $\mathcal{F}$. Hence, at this moment, the charges possess the Coulomb energy $E(t=0) = \mathcal{F}\mathcal{R}$ (Eq2). If for the time $\tau_w$ (at the end of working stroke), the distance between the centers of spherules increases and becomes $\mathcal{R}+D$, their Coulomb energy decreases in accordance with (Eq2) to

$$Q \equiv E\left(t = \tau_w\right) = \frac{\mathcal{F}\mathcal{R}^2}{\mathcal{R} + D}\,. \tag{Eq4}$$

This Coulomb energy will be brought away by the charges escaped from the cylinder and dissipate. The mechanical work performed by the Coulomb forces is

$$W \equiv E\left(t = 0\right) - E\left(t = \tau_w\right) = \frac{\mathcal{F}\mathcal{R} D}{\mathcal{R} + D}\,, \tag{Eq5}$$

so that the efficiency of the system without friction depends only on the relationship between $\mathcal{R}$ and $D$ and is equal to

$$\eta_0 \equiv \frac{W}{\mathcal{F}\mathcal{R}} = \frac{D}{\mathcal{R} + D}\,. \tag{Eq6}$$

Then, using the lever (see Fig. 2) or some other means, the work (Eq5) should be converted into the work on displacement of the rope. The average force applied to the rope from the lever is conventionally equal to the work performed divided by the rope displacement. The force averaged over the *working stroke duration*, is equal to

$$\langle f \rangle_{ws} \equiv \frac{W}{S} = \frac{1}{\lambda D}\int_0^D \mathcal{F}\left(\mathcal{R} + x\right)dx = \frac{\mathcal{F}}{\lambda}\frac{\mathcal{R}}{\mathcal{R} + D} \tag{Eq7}$$

and does not depend on the rope velocity. This average force becomes equal to the maximal Coulomb force $\mathcal{F}$ multiplied by two factors. The existence of the lever is the reason for the first factor, $1/\lambda$, to appear. The second factor, $\mathcal{R}/(\mathcal{R}+D)$ appears in the course of averaging of the repulsive force between $q_1$ and $q_2$, which diminishes from $\mathscr{F}$ up to $\mathscr{F}[\mathscr{R}/(\mathscr{R}+D)]^2$ from the beginning to the end of the working stroke.

The most interesting, however, is the force value $\overline{f}$ averaged *over the period*, which already is the function of the rope velocity, rather than $\langle f \rangle_{ws}$. As long as the velocity of the rope relative to the cylinder movement is changeless and equal to $v$ and the period of back and forth motion of the engine is $T$, this average value is equal to

$$\overline{f}(v) \equiv \frac{W}{vT} = \frac{\mathscr{F}}{vT}\frac{\mathscr{R}D}{\mathscr{R}+D} = \frac{\tau_w(v)}{\tau_r + \tau_w(v)} \cdot \langle f \rangle_{ws}. \qquad \text{(Eq8)}$$

I should remind that we suppose here that the transmission gear functions without losses. In this case the connection between the force $\overline{f}(v)$ and the rate of energy dissipation $Q(v)/T$ looks as follows:

$$\frac{Q(v)}{T} = \frac{\mathscr{R}}{D}v\overline{f}(v), \qquad \text{(Eq9)}$$

i.e. the rate of heat production is proportional to the average mechanical power $v\overline{f}(v)$. It is evident that this result is a logical consequence of the proportion $W/Q = D/R$ (see (Eq4) and (Eq5)) if one takes into account the relation between $W$ and $\overline{f}$ (Eq8).

To move further, we have to take into account that $\tau_w$ depends on the rope velocity, whereas $\tau_r$ does not. The connection between the distance $D$ at which the spherule 2 is displaced during the working stroke, the corresponding time span $\tau_w$ and the rope velocity $v$ appears from (Eq3):

$$\tau_w(v) = \frac{\lambda D}{v}. \qquad \text{(Eq10)}$$

Now it is easily seen from (Eq8) that if $\tau_r$ does not depend on $v$, these model yields a hyperbolic form of the force-velocity curve:

$$\overline{f}(v) = \frac{\mathscr{F}}{\lambda D + v\tau_r}\frac{\mathscr{R}D}{\mathscr{R}+D} = \frac{\langle f \rangle_{ws}}{1 + \dfrac{v\tau_r}{\lambda D}}, \quad \langle f \rangle_{ws} = \overline{f}(v=0).. \qquad \text{(Eq11)}$$

It is clear that as the velocity diminishes, the working stroke takes still the greater and greater part of the period, so the average values $\langle f \rangle_{ws}$ (Eq7) and $\overline{f}(v)$ (Eq8) just coincide if $v$=0.

We can consider now the existence of some velocity-independent hindering force $F_h$, which acts in the transmission gear, and subtract it from the right hand part of (Eq11). As a result we obtain:

$$F(v) \equiv \overline{f}(v) - F_h \quad \Rightarrow \quad \frac{v}{v_{max}} = \frac{1 - \dfrac{F(v)}{F_{max}}}{1 + \dfrac{F_{max}}{F_h}\dfrac{F(v)}{F_{max}}},$$

$$F_{max} \equiv \langle f \rangle_{ws} - F_h = \frac{\mathscr{F}\mathscr{R}}{\lambda(\mathscr{R}+D)} - F_h, \qquad v_{max} \equiv \frac{\lambda D}{\tau_r}\frac{F_{max}}{F_h}. \tag{Eq12}$$

It is just the standard normalized form of the Hill equation (Hill, 1938: Woledge, Curtin and Homsher, 1985) that describes the experimentally observed relationship between the stress produced by a muscle and its contraction velocity. It should be noted also that the hyperbolic-like form of the $F(v)$ dependence (Eq12) (or, which is almost the same, of the $\overline{f}(v)$ dependence (Eq11), see Discussion also) is a direct sequence of two conditions:

1. Recharging time $\tau_r$ does not depend on the rope velocity.

2. The working stroke duration $\tau_w$ is inversely proportional to the rope velocity (Eq10).
 These two conditions are intrinsic in the model and do not depend on a concrete type of the transmission gear. No other speculative assumptions are needed to obtain the Hill equation in the framework of the model.

Two important consequences are worth noting here that can be easily derived from these formulae. The first one can be obtained from (Eq10), (Eq11) and the condition $F(v_{max}) = 0$:

$$\frac{\tau_r}{\tau_w(v_{max})} = \frac{F_{max}}{F_h}. \tag{Eq13}$$

It means that the well-known factor $F_{max}/F_h$ regulating the curvature of the hyperbola (Eq12) is controlled by the recharging and working stroke times in this model. The other important formula describes the velocity dependence of efficiency. In the case $F_h$=$const$ it has the form

$$\eta(v) = \eta_0 \frac{\overline{f}(v) - F_h}{\overline{f}(v)} = \eta_0 \left(1 - \frac{F_h}{\langle f \rangle_{ws}}\right)\left(1 - \frac{v}{v_{max}}\right). \tag{Eq14}$$

One can easily see that the dependence is linear in $v$ and, in addition, even the maximal value $\eta(v=0)$ is less than $\eta_0$ (Eq6). The reason is the additional heat production connected with the appearance of hindering force.

## IV. ISOVELOCITY MUSCLE CONTRACTION

Quite a simple model discussed in the previous section properly reproduces, as I believe, the cross-bridge performance. To affirm the truth of the statement, let me cite (Geeves and Holmes, 1999, page 703). "Opening the switch-2 region destroys the γ-phosphate-binding pocket and … would appear to facilitate γ-phosphate release (a "back door enzyme"). …. The movement of the switch-2 in the closed form has other more far-ranging consequences, namely the rotation of the converter domain through about 60°. …. The end of the lever arm has moved through 11 nm along the actin helix axis between open and closed, which is about the expected magnitude of the power stroke. This large change is driven through molecular cogs and gears by a small (0.5 nm) change in the active site. Therefore, it now seems rather likely that the myosin power stroke works by switching between these two conformations."

To see in this description the model discussed in the previous section, it is enough to suppose that switch-2 moves under the action of $P_i$ which is repelled from $ADP$ by Coulomb force. In this case we find the following parameter values: $D \approx 0.5$ nm, $S \approx 10$ nm, $\lambda \approx 20$. It is known also (Huxley, 2000;Woledge, Curtin and Homsher, 1985, section 1.3) that the force exerted by a single cross-bridge in isometric contraction (i.e. $F_{max}$ in the model) is about 4 pN, and the value of the free energy change upon the $ATP$ splitting is about $\Delta G_{ATP} \approx 50$ kJ/mol $\approx 8 \cdot 10^{-20}$ J ($ATP$ molecule)$^{-1}$. The difference between this free energy and the Coulomb energy of a single $ATP$ molecule immediately after hydrolysis appears because of the necessity to thermolize $ADP$ and $P_i$ (the total number of their degrees of freedom is larger than that of $ATP$ molecule). However, this difference in energy is on the order of magnitude of $k_B T \approx 4 \cdot 10^{-21}$ J where $T \approx 300$ K is the absolute temperature, and in the simplest approximation can be neglected in comparison with $\Delta G_{ATP}$.

To proceed, we should now write for the $ATP$ molecule cleavage process the equation analogous to (Eq2). As is stated above, in the milieu, the $ATP$ molecule keeps four extra electrons. Each of the two closest to adenosine residuals of the phosphoric acid bears solely one electron, whereas the third residual has double electron charge (Bendall, 1969; Bohinski, 1983).

Suppose that *ADP* and $P_i$ at a moment $t = 0$ just after the hydrolysis keep the charges $n_1 e$ and $n_2 e$, respectively ($e \approx 1.6 \cdot 10^{-19}$ C is the charge of electron), and the dielectric permeability of the milieu is $\varepsilon$. Suppose then that after the hydrolysis *ADP* and $P_i$ are localized in the *ATP*-binding center at a distance $\mathcal{R}$ from each other, the Coulomb force acting on them is $\mathcal{F}$ and the system possesses energy $\mathcal{E}$. Hence, at this moment we have for Coulomb force and energy the formulae

$$\mathcal{F} = 9 \cdot 10^9 \frac{n_1 n_2}{\varepsilon} \frac{2.56 \cdot 10^{-38}}{\mathcal{R}^2} \approx 2.3 \cdot 10^{-28} \frac{(n_1 n_2 / \varepsilon)}{\mathcal{R}^2} \text{ N}, \quad \mathcal{E} = \mathcal{F}\mathcal{R} \text{ J}. \quad \text{(Eq15)}$$

Starting from (Eq15) we can reproduce all the formulae (Eq4)÷(Eq14), but to find the value of any physical magnitude in these equations we anticipatorily have to estimate the values of two more experimental parameters. Considering the hindering force as independent of contraction velocity, we can for example use the standard quantities characteristic of force-velocity curve $F_{max} / F_h \approx 4$ and $v_{max} \approx 1.5 \, \mu\text{m/s}$ (Woledge, Curtin and Homsher, 1985, Table 2.II). Using these parameter values, we, for the reasons discussed in the following section, apply for $F_{max}$ the formula

$$F_{max} = \frac{1}{\lambda} \mathcal{F} - F_h, \quad \text{(Eq16)}$$

i.e. drop the multiplier $\mathcal{R}/(\mathcal{R}+D)$ in the formula (Eq12). Hence, alongside with (Eq15) we use equations

$$\begin{cases} \mathcal{F} \approx \lambda F_{max} \left(1 + \dfrac{F_h}{F_{max}}\right) \\ \mathcal{E} \approx \Delta G_{ATP} \\ \tau_r \approx \dfrac{F_{max\,h}}{F_h} \dfrac{\lambda D}{v_{max}} \end{cases} \quad \text{where} \quad \begin{bmatrix} \lambda \approx 20 \\ F_{max} \approx 4 \cdot 10^{-12} \text{N} \\ \dfrac{F_{max}}{F_h} \approx 4 \\ \Delta G_{ATP} \approx 8 \cdot 10^{-20} \text{J} \\ D \approx .5 \cdot 10^{-9} \text{m} \\ v_{max} \approx 1.5 \cdot 10^{-6} \text{m s}^{-1} \end{bmatrix} \quad \text{(Eq17)}$$

Herefrom, we find

$$\begin{cases} 2.3 \cdot 10^{-28} \dfrac{(n_1 n_2 / \varepsilon)}{\mathcal{R}^2} \approx 20 \cdot 1.25 \cdot 4 \cdot 10^{-12} \\ 2.3 \cdot 10^{-28} \dfrac{(n_1 n_2 / \varepsilon)}{\mathcal{R}} \approx 8 \cdot 10^{-20} \\ \tau_r \approx 4 \cdot \dfrac{20 \cdot 0.5 \cdot 10^{-9}}{1.5 \cdot 10^{-6}} \end{cases} \quad \Rightarrow \quad \begin{cases} n_1 n_2 / \varepsilon \approx .28 \\ \mathcal{R} \approx .8 \text{ nm} \\ \tau_r \approx 2.7 \cdot 10^{-2} \text{ s} \end{cases} \quad \text{(Eq18)}$$

The numerical values obtained are quite reasonable by the order of magnitude. Indeed, supposing that any of the two *ATP* remainders, *ADP* and $P_i$ keeps two extra electrons ($n_1=n_2=2$), we have $\mathcal{E} \approx 15$ (for water $\mathcal{E} \approx 80$), while supposing that the milieu does not leak in the *ATP*-binding pocket ($\mathcal{E}=1$), we gain the remainder charges $n_1 = n_2 \approx \frac{1}{2}$ that are slightly lower that in the previous case. The latter is more likely (see Discussion) and the divisibility of the obtained *n* values is traceable to the fact that the real distribution of charge density inside *ATP*-binding pocket naturally does not come to two point charges accounted for in the model.

The value $\mathcal{R} \approx .8$ nm is quite reasonable as well. Inserting this value and the value $D \approx 0.5$ nm into (Eq6) we find the cross-bridge efficiency $\eta_0 \approx 40\%$ and the maximal efficiency of muscle (Eq14) $\eta(v=0) \approx 30\%$. It should be commented here that using the formula (Eq12) for $F_{max}$ which includes the multiplier $\mathcal{R}/(\mathcal{R}+D)$ one obtains the noticeably less value of $\mathcal{R}$ and hence considerably larger values of $\mathcal{F}$ and $\eta$. Nevertheless, for the reasons discussed in the following section we chose the estimation (Eq16), (Eq18).

Further, the period of cross-bridge back and forth motion $T(v) = \tau_r + \lambda D / v$ is equal to 33 ms (the frequency 30 Hz) if v=$v_{max}$ and increases up to 93 ms (the frequency $\approx 11$ Hz) upon the tenfold decrease of velocity. Finally, it should be noted that the formula for the maximal velocity of muscle contraction $v_{max}$, in which $\tau_r$ stands in the denominator at least does not contradict experiments. The recovering time $\tau_r$ is summed of the times of duration of several processes, one of which is the process of *ADP* release after completion of power stroke. Therefore, the observed experimental linear relation between $v_{max}$ and *ADP* release rate (Weiss *et al* 2001) agrees well with (Eq12) (see the next section also).

It is quite pertinent now to rewrite formulae of the previous section in the more convenient and habitual form. For example, taking into account the heat production connected with the hindering force existence, we obtain from (Eq9)

$$\frac{H(v)}{F_{max}} = v \cdot \frac{F_h}{F_{max}} \cdot \left\{ 1 + \frac{\mathcal{R}^2}{D(\mathcal{R}+D)} \cdot \frac{1 + \dfrac{F_{max}}{F_h}}{1 + \dfrac{F_{max}}{F_h} \cdot \dfrac{v}{v_{max}}} \right\}. \qquad \text{(Eq19)}$$

Qualitatively, the dependence (Eq19) quite correctly reproduces the specific form of the experimentally observed curves of heat production rate vs. velocity $H(v)$; see (Woledge, Curtin and Homsher, 1985, Fig. 4.13 and the references in the capture to this figure). Nevertheless, the quantitative difference here is principal. First, a muscle produces some amount of heat during isometric contraction while (Eq19) gives $H(v=0) = 0$. Secondly, using (Eq19) and the parameter values (Eq17), (Eq18) (the experiments mentioned above and the set of parameters are related to the same sartorius of *R. temporaria*) one obtains

$$\frac{H(v) - H(v=0)}{v \cdot F_{max}} \approx \begin{cases} 1.5 & \text{if} \quad v \to 0 \\ .5 & \text{if} \quad v = v_{max} \end{cases} \qquad \text{(Eq20)}$$

which is several times as high as all the experimental values. Any attempt to agree these contrarieties in the framework of the model is apparently foredoomed to failure, so we have a significant discrepancy between the model prediction and the experimentally observed rate of heat production in the low-velocity range.

## V. LOW-VELOCITY RANGE

It is easy to assure ourselves that this discrepancy is not single. For example, according to (Eq14) the efficiency does not vanish in the case of isometric contraction though the mechanical work is absent in this case. Moreover, even the real force-velocity curve does not obey Hill equation in the low-velocity range; see (Edman *et al*, 1976; Edman, 1988). Hence, the model obviously fails in this range.

The problem is traceable to the assumption, which is not groundless but not wholly true, used in the model. We have supposed that the only way for *ATP* remainders *ADP* and $P_i$ to release from *ATP*-binding center is the opening the backdoor that results in the relative displacement of actin and myosin filaments. The duration of this process in quasiisometric case ($v \to 0$) should be infinite and, consequently, the rate of heat production and the mechanical power vanish. Nevertheless, the ratio of these quantities does not vanish and neither does the efficiency.

In spite of these predictions, in the course of isometric contraction of a real muscle the displacement of filaments is absent while the cross-bridges move on attaching to and detaching from actin with the period of the order of 0.5 s (Woledge, Curtin and Homsher, 1985). The efficiency in this case is evidently equal to zero and the distance between *ADP* and $P_i$ is the same over the working stroke. So, the force that the cross-bridge exerts to actin should not include the multiplayer $\mathscr{R}/(\mathscr{R}+D)$ which appears in (Eq7) after averaging on the distance (compare (Eq12) and (Eq16)).

To make the situation clear, we apparently must allow for the fact that the system under discussion (*S1*) is characterized by a size of 10 nm and captures and splits a single molecule. Hence, fluctuations should play quite an important role in the cross-bridge performance, and the release of the *ATP* hydrolysis products in the isometric (or quasiisometric) contraction case is very likely to be the result of some stochastic process. Therefore, it seems natural to suppose that the actomyosin complex possesses an intrinsic lifetime and its duration depends on certain stochastic processes. Knowing nothing about the nature of these processes, we can introduce a single parameter to describe them, which we denote $\tau_0$. In the framework of the model from Fig.2 it means that if at the moment $\tau_0$ the piston, which was displaced by the distance

$$d(v) \equiv \frac{1}{\lambda} v \tau_0, \qquad (Eq21)$$

did not leave the cylinder yet (i.e. $d(v) < D$), the working stroke is terminated, the charges are released and the energy they possess at the moment converts to heat. It is clear that if the inequality $\tau_0 > \tau_w = \lambda D / v$ holds true, this stochastic mechanism does not have enough time to dissociate the actomyosin complex, and the mechanism described above works. On the contrary, if the opposite inequality holds true, the stochastic mechanism works, and the change of mechanisms occurs if the velocity becomes equal to

$$v_{ch} = \frac{\lambda D}{\tau_0}. \qquad (Eq22)$$

At last, one can thing that the averaged time interval $\tau_r$ between the moments of the hydrolysis products releasing and beginning of the next working stroke (recovering time) should depend on the *ATP* concentration and some other chemical parameters rather than on the contraction velocity. In this case $\tau_r$ is independent of contraction velocity, and the period of breaking and reforming of a cross-bridge $T$ depends on $v$ for high contraction velocities and does not depend for low ones.

It is clear that when the duration of working stroke does not depend on velocity, formulas (Eq4)÷(Eq8) are valid as well providing the only modification being made – substitution $D \rightarrow d(v)$ (Eq21):

$$\langle f \rangle_{ws} = \frac{\mathscr{F}}{\lambda} \frac{\mathscr{R}}{\mathscr{R} + d(v)} = \frac{\mathscr{F}}{\lambda} \frac{1}{1 + \dfrac{\tau_0 v}{\lambda \mathscr{R}}}; \quad \overline{f}(v) = \frac{\tau_0}{\tau_0 + \tau_r} \langle f \rangle_{ws} \qquad (Eq23)$$

It is easily seen from (Eq7), (Eq8) and (Eq23) that the two mechanisms yield the same hyperbolic-shape form of the force-velocity curve, but their parameters and the reasons that condition this very dependence are entirely different. At a high contraction velocity, $v > v_{ch}$, the working stroke duration $\tau_w$ depends on velocity and, correspondingly, the average number of cross-bridges attached to actin at a moment, which is proportional to $\tau_w / (\tau_r + \tau_w)$, depends on velocity as well. Therefore, in this case one can also say that the force generated by a single cross-bridge $\langle f \rangle_{ws}$ does not depend on velocity while the fraction of cross-bridges, attached to actin at a moment, falls as the velocity rises. The stochastic mechanism, if it actually works at low velocities $v < v_{ch}$, keeps the proportion of attached and detached cross-bridges unchanged. It is the average force acting during working stroke $\langle f \rangle_{ws}$ that is now dependent on velocity, therefore the Hill equation in this case is a direct consequence of the Coulomb law, see (Eq23).

Thus, introduction in consideration of the stochastic mechanism of interrupting the working stroke seemingly allows one to lift most problems arising in the model at low contraction velocities. The heat release becomes weakly dependent on the velocity now so that the ratio of cross-bridge capacity to the heat released and hence, efficiency vanishes at $v \rightarrow 0$. Moreover, this puts forward an explanation of the experimentally observed deviation of the force-velocity dependence from a single Hill hyperbola in the range of low velocities. The model predicts now that this curve must consist of two hyperbolas converging in the point $v = v_{ch}$ (Eq22). However, assuming that $\tau_0$ magnitude is of the order of some tenth of second (Woledge, Curtin and Homsher, 1985), we obtain that $v_{ch} \approx 25$ nm s$^{-1}$, which is lower than the experimentally observed (Edman 1988) value $v_{ch} \approx v_{max}/10 \approx 150$ nm s$^{-1}$ by approximately the order of magnitude. Besides, the velocity of heat release at $v < v_{ch}$ now turns out about constant and this result is inconsistent with experiment (Fenn effect).

This plausibly means that even in the low-velocity region the period of cross-bridge oscillations has to decrease with increasing velocity. To clarify the reason for that it should be understood what kind of fluctuations causes the release of *ADP* and $P_i$ from the *ATP*-binding pocket at low velocities of muscle contraction. However, because of the lack of distinct experimental data we can so far put forward solely more or less grounded guesses.

First of all, it is difficult to expect that rather a massive inorganic phosphate (and all the more *ADP*) can tunnel through the pocket wall so that their release must be connected with the backdoor opening. Immediately after hydrolysis it is acted on by a significant force from $P_i$ but because of counteraction of stretched cross-bridge neck, the backdoor remains close. To make it open a little, it is apparently necessary to either additionally stretch the cross-bridge neck or temporarily get the myosin head detached from actin filament.

The latter process seems significantly more probable for some reasons. In particular, the binding energy of the actin-myosin complex should be relatively small in comparison with $\Delta G_{ATP} \approx 20 k_B T$, which results in a relatively high probability of such fluctuations. If they actually arise and myosin head gets detached from the actin filament, the backdoor must get open at this moment under the inorganic phosphate pressure (and myosin head, consequently, move aside) so that $P_i$ gets a chance to slip outside.

At this step, uncertainty arises again. If *ADP* leaves *ATP*-binding pocket together with $P_i$, we return to (Eq23). However, a far more massive molecule *ADP* may fail to free itself together with $P_i$. In this case, since the affinity of myosin head for actin remains high, after repeated attachment of myosin head to actin in a new position, the *ADP* molecule must stay inside the *ATP*-binding pocket. The time of staying there will depend on the state of backdoor or, what is the same, on the position of converter domain. If backdoor is closed, *ADP* has to stay inside until the next more significant fluctuation comes, or, if the muscle contraction proceeds with a high enough velocity, until the backdoor opens owing to the filament sliding. However, in this case the internal pressure on the backdoor is absent; therefore it has to open only after the neck stretching in opposite direction. Taking into account that the length of the flexible hinge region is on the order of 40 nm, we obtain the filament displacement of about 80 nm, which just results in a correct value of $v_{ch}$.

To describe this mechanism, simple and quite evident changes must be introduced in the model. However, this can be considered reasonable only if precarious grounds, which serve as a basis for this reasoning, will be experimentally supported. It is interesting to note, however, that from this point of view different experimentally observed phenomena – stereo-specifically and non-stereo-specifically bound myosin heads (the roll and lock mechanism (Ferenczi *et al*, 2005)), strain-sensitive *ADP* release mechanism (Nyitraiy and Geeves,. 2004) and values of the myosin head step, changing in the broad limits (Yanagida *et a,l* 2000), may turn out merely different manifestations of one and the same process. Nevertheless, the dependence of recovering time $\tau_r$ on velocity may arise because of different cooperation effects as well, say, those related to calcium influence (Gordon, Homsher and Regner, 2000; Katsnelson and Markhasin, 1996).

## VI. DISCUSSION

Despite the significant problems arising in the low-velocity region, the advantages of the model suggested are evident. In general, the model correctly describes the dependences of the force, efficiency, and heat release intensity on the contraction velocity. It is of no difficulty either to apply the model to description of transition processes, at least, upon abrupt *decreasing* the muscle length. It is clear that with such filament displacement at a distance exceeding $\lambda D$ (in counting of half the sarcomere) the backdoors of all the cross-bridges turn out open and they detach from actin so that the muscle stops generating a force. At a smaller displacement, this happens only with a some (easily calculated) part of cross-bridges, but in any case, the subsequent transition processes of recovering the muscle strength will be connected with the recovering of actomyosin complexes.

Here, it should be noted one essential circumstance. Though the model treats only a single cross-bridge, this by no means purports that all the cross-bridges work independently. The matter is that all of them are attached to the same actin filament and move with the same velocity, i.e., there is a *kinematical* restraint between them. Apparently, there are no grounds to think that along with this restraint there is another *direct* interaction between different cross-bridges. This, in particular, means that stresses arisen in the neck of each cross-bridge may be only related either to interactions *inside* the cross-bridge or muscle stretching *in toto*. Therefore, actin filament displacement is the only means for a single cross-bridge to feel the change in the external load on the muscle. It is just the decrease in the velocity of contraction, which necessarily arises upon the increase in the load on the muscle, that causes a *synchronic* increase of the force generated by each cross-bridge, i.e., it causes any load to be raised by a muscle uniformly. Such mechanism of restraining arises because the contraction velocity becomes the principal parameter of the model now. This mechanism is radically different from the mechanism of the influence of stresses on the cross-bridge dynamics, which was suggested by (Huxley and Simmons, 1971) and used many times in other works, see, for example (Duke 1999). In connection with this, I permit myself one more citation (A.F. Huxley 2000). "The fact that the response to length change is composed of first-order delays while that to load change is oscillatory implies that the molecular events are directly affected by longitudinal displacement of the filaments rather than by the tension in them."

The critically important result of the model is also the quite reasonable order of magnitude of numerical values of the theory parameters obtained in (Eq18). This coincidence is the strong argument testifying to the validity of the model but it is not the quietus. The most important evidence to the fructuousness of model, along with the level of coincidence of its predictions with the experimental data, is the validity and *modesty* of the assumptions underlying it. It is just the analysis of the assumptions underlying the model that is the subject matter of discussion below.

As to the Coulomb interaction, the first and main assumption, i.e., the basic postulate of this work (see section II) is hardly doubtful. Much more vulnerable is the involvement of the simplest formulas (Eq2) and (Eq15). Actually, however, the model deals only with the dependence of energy on distance of $1/r$ type that is quite stable and does not change, say, with allowance for quantum effects. Noticeable deviations from the $1/r$ law for the charges interacting in a dielectric medium can arise only in the case when the shape of the charges essentially differs from a spherical and their sizes are large compared to the distance between them. Otherwise, the force acting on the charges only decreases in magnitude because of the dielectric permeability of the medium $\varepsilon$, and that is made allowance for this decrease in the model.

The situation changes radically if the charges interact in conductive medium. The Debay-Hückel mechanism of screening by free electrons and ions appears now, and the dependence of the energy of interacting charges $E$ on distance $r$ takes the form

$$E(r) \sim \frac{1}{r} \exp\left(-\frac{r}{l}\right) \qquad (Eq24)$$

in this case (Landau and Lifshits, 1964). In the normal cell-milieu the order of magnitude of the Debay-Hückel radius $l \approx 1$ nm, and it seems as if one has to use (Eq24) rather than (Eq15) in the model suggested above. Alteration of this kind cannot change the dependences of force, efficiency and the rate of heat production on velocity, which is connected with $\tau_r$ and $\tau_w$ dependences on $v$. At the same time, the values of the force and efficiency decrease the more significantly, the less the screening radius $l$ is, and for $l \lesssim 1$ nm the model works much worse.

I believe that (Eq15) rather than (Eq24) have to be the base for the model. Attempting to back these words with proof, one can find quite a reasonable explanation of the cross-bridge working cycle, which, at first glance, looks unwarrantedly complicated. Indeed, the *ATP* molecule attached to the myosin *ATP*-binding center first is blocked by the special door in *ATP*phase-pocket, and its splitting takes place. Then, the remainders are let go outside owing to the motion of the door, which is connected with the complex system of levers. This principle of cross-bridge performance is strikingly alike the operation of gas-engine, and two following questions immediately appear here.

The former one is what the role is played by the walls of *ATP*phase-pocket (compare the walls of gas-engine cylinder or the trunk), i.e. why the *ATP*-molecule has to be isolated before its splitting. The answer depends on the porosity of the *ATP*phase-pocket's walls. If these walls are impenetrable to free ions, the natural assumption is that it just this pocket that blocks the Debay-Hückel screening, so that (Eq15) rather than (Eq24) is the base for the model. It should be noted also that the *ATP* isolation in the *ATP*phase pocket is evidently related to the "tearing off the charged fur" in which an *ATP* molecule is dressed in the milieu. Therefore, the localization of the molecule in the pocket requires a work to be performed (compare the compression of air and fuel mixture in a cylinder of gas-engine). It implies in its turn that cross-bridge conserves some energy store even *after* completing the working stroke and thus, we return again to some "myosin state in which it can do work" (see section II). Nevertheless, this is not the total energy developed in the course of *ATP* cleavage; only some part of this energy should be stored.

The latter question is which way the door of *ATP*-binding-pocket functions? On the one hand, the door, just as the walls, shields *ATP* of the milieu, and this function is clear. Nevertheless, on the other hand, it blocks *ATP* in the pocket and the cornerstone question raises here: what is the essence of this blocking or, in other words, why does the door open and what does the system of levers connected to it mean?

One can think that when the cross-bridge changes its form in the course of working stroke, the levers pull and move the door, opening it and releasing *ADP* and $P_i$. This notwithstanding, in the framework of the model suggested we have to think that inorganic phosphate group, broken away from *ADP* and pushing off it, itself pushes the door, this force propels the levers, and the cross-bridge form changes - power-stroke appears. From this point of view, the role of the levers is quite obvious as well. The Coulomb force sharply decreases with distance, and its value is close to the initial one only on the distance of the order of 1 nm. To raise the pace up to 10 nm the levers are needed, so the levers play the role analogous to the gear-box role.

In conclusion I would like to underline once more that in the framework of the model the Hill equation appearance is the effect of the three basal presumptions:

1. The energy of interaction between *ADP* and $P_i$ depends on distance between them only and hence, the efficiency *of a single cross-bridge* (Eq6) does not depend on contraction velocity.

2. $\tau_r$ and $(v\tau_w)$ do not depend on $v$ too.

3. Some independent of $v$ "hindering" force exists.

In addition to first two suppositions having been discussed in section III, here it is necessary to particularly pick out the assumption on the existence of a certain hindering force. The matter is that in the mechanical model in Fig.2 the existence of friction force is quite natural, whereas in a muscle the reason for its emergence is by no means evident. Especially, this problem is complicated in view of the results of the work (Uyeda *et al* 1996) where experiments were conducted on mutant myosins that have different neck lengths. They unambiguously evidence that even in unloaded state the angular velocity of rotation of a single *S1* neck is limited, which purports that the hindering force arises inside each individual myosin head rather that in the whole muscle.

**ACKNOLEDGEMENTS**

I am grateful to Dr. M. L. Shur who has attracted my attention to this problem and to Dr. S.Y. Bershitsky for his criticism and kindly permission to learn his exemplar of the book (Woledge, Curtin and Homsher, 1985).